\documentclass[aps,prd,twocolumn,letterpaper,showpacs,nofootinbib,floatfix]{revtex4}
\usepackage{bm}
\usepackage{graphicx}
\usepackage{latexsym}
\usepackage{amsmath}
\usepackage{amsfonts}
\usepackage{amssymb}
\usepackage[pdftex,bookmarks=true,bookmarksopen=true,bookmarksnumbered=true,bookmarksopenlevel=3]{hyperref}
\hypersetup{
    bookmarks=true,     
    unicode=false,          
    pdftoolbar=true,        
    pdfmenubar=true,        
    pdffitwindow=false,     
    pdfstartview={FitH},    
    pdftitle={DRAFT},    
    pdfauthor={Orlando Panella},     
    pdfsubject={UED level-1 bound states},   
    pdfcreator={latex},   
    pdfproducer={TEXSHOP}, 
    pdfkeywords={UED, Bound states }, 
    pdfnewwindow=true,      
    colorlinks=true,       
    linkcolor=red,          
    citecolor=blue,        
    filecolor=magenta,      
    urlcolor=cyan           
}

\newcommand{\beq}{\begin{equation}}
\newcommand{\eeq}{\end{equation}}
\newcommand{\beqy}{\begin{eqnarray}}
\newcommand{\eeqy}{\end{eqnarray}}

\newcommand{\as}{\alpha_s}

\begin{document}

\title{Threshold production of meta-stable  bound states of Kaluza Klein
excitations in Universal Extra Dimensions}

\author{ {\textsc{N.~Fabiano and O.~Panella}}}
\affiliation{Istituto Nazionale di Fisica Nucleare,
Sezione di Perugia, Via A. Pascoli I-06123, Perugia, Italy}

\date{\today}

\begin{abstract}
We study the formation and detection at the next linear $e^+e^-$ collider of bound states of level-1 quark Kaluza-Klein
excitations ${\cal B}_{KK}$ within a scenario of universal
extra-dimensions (UED).  The interactions of such Kaluza-Klein excitations are modeled  by an $\alpha_s $ driven Coulomb potential.  In order to obtain the threshold
cross-section, we employ the Green function method which is
known to properly describe the peaks below threshold and to
yield a net increase in the continuum region (above
threshold) relative to the naive Born cross-section. We study such effect at different values of the scale ($R^{-1}$) of the extra-dimensions with an explicit calculation of the mass spectrum as given by radiative corrections. The
overall effect is roughly 2.7 at $R^{-1}=300$ GeV and goes down to $2.2$ at $R^{-1}=1000$ GeV and a relatively large
number of events is expected from $N_{events} \approx 2.5\times 10^4$ at $R^{-1}=300$ GeV down to $N_{events} \approx 10^3$ at $R^{-1}=1000$ GeV 
at the anticipated annual
integrated luminosity of $L_0= 100$ fb$^{-1}$. We finally discuss
some potentially observable signatures such as the multilepton channels 
$2j +2\ell +E\!\!\!\! /$\, and  $2j +4\ell +E\!\!\!\! /$ for which we estimate statistical significance $\gtrsim 2$ for $R^{-1}$ up to  $600 \sim 700$ GeV.  
\end{abstract}

\pacs{12.60.-i, 11.10.St, 14.80.-j}

\maketitle
\section{Introduction}
\label{sec:intro}

It is well known that as early as 1921 Theodore Kaluza 
proposed a theory that was intended to unify gravity and
electromagnetism by considering a space-time with one
extra \emph{space-like} dimension~\cite{Kaluza:1921tu}. A
few years later Oscar Klein proposed that the extra space
dimension (the fifth dimension) is in reality compactified
around a circle of very small radius~\cite{Klein:1926tv}.
These revolutionary ideas have thereafter been ignored for
quite some time. However recent  developments in the field
of string theory have suggested again the possibility that
the number of space time dimensions is actually different
from $D=4$ (indeed string theory models require $D= 11 $,
i.e. seven additional dimensions). In 1990 it was
realized~\cite{Antoniadis:1990ew}  that string theory
motivates scenarios in which the size of the extra
dimensions could be  as large as $R\approx 10^{-17}$
cm (corresponding roughly to electro-weak  energy
scale ($\approx$ TeV) contrary to naive expectations  which
relate them to  a scale of the order of the Planck length
$L_P\approx 10^{-33}$ cm (corresponding to the
Planck mass $M_P=\sqrt{\hbar c/G} \approx 10^{19}$ GeV).
See also~\cite{Antoniadis:1998ig}.

Subsequently two approaches have been developed to discuss the observable
effects of these, as yet, hypothetical extra dimensions. One possibility is to
assume that the extra space-like dimensions are flat and compactified to a
``small'' radius. This is the so called ADD model~\cite{ArkaniHamed:1998rs}
where
only the gravitational interaction is assumed to propagate in the
extra-dimension. A second possibility is contemplated in the Randall-Sundrum
type of models where the extra dimensions do have curvature and are embedded in
a warped geometry~\cite{Randall:1999ee,Randall:1999vf}.

Universal extra-dimensional models were introduced in
ref.~\cite{Appelquist:2000nn} and are characterized, as opposed to the ADD
model, by the
fact that all particles of the Standard Model (SM) are
allowed to propagate in the (flat) extra space dimensions,
the so called \emph{bulk}. Here to each SM particle $X^{(0)}$
corresponds in this model a tower of Kaluza-Klein states
$X^{(n)}$ (KK-excitations),  whose masses are related to
the size of the compact extra dimension introduced and the
mass of the SM particle via the relation $m^2_{X^{(n)}}
\approx m^2_{X^{(0)}}+n^2/R^2$. An important aspect of the
UED model is that it provides a viable candidate to the
Cold Dark Matter. This would be the lightest KK particle
(LKP) which typically is the level 1 photon. Many aspects
of the phenomenology of these KK excitations have been
discussed in the literature. For reviews see
ref.~\cite{Hooper:2007qk,Macesanu:2005jx,Rizzo:2010zf,Cheng:2010pt}. In particular KK production has been considered both at the Cern large hadron collider
(LHC) and at the next linear collider (ILC). Direct
searches of KK level excitations at collider experiments
give a current bound on the scale of the extra-dimension of the order
$R^{-1} \gtrsim 300$ GeV. See for example
ref.~\cite{PDBook}. At the Fermilab Tevatron it will be possible to test
compactification scales up to $R^{-1} \sim 500$ GeV at least within some
particular scenario~\cite{Macesanu:2002db,Macesanu:2002ew,Macesanu:2002hg}.

Lower bounds on the compactification radius arise also from analysis of electro-weak precision measurements performed at the $Z$ pole (LEP II). An important feature of these type of constraints is their  dependence on the Higgs mass. A recent refined analysis~\cite{Gogoladze:2006br} taking
into account sub-leading contributions from the new physics  as well as two-loop corrections to the standard model $\rho$ parameter finds that
$R^{-1} \gtrsim 600$ GeV for a light Higgs mass ($m_H= 115$ GeV) and a top quark  mass $m_t=173$ GeV at 90\% confidence level (C.L.). Only assuming a larger value of the Higgs mass the bound is considerably weakened down to $R^{-1} \gtrsim 300$ GeV for $m_H= 600$ GeV, thus keeping the model within the reach of the Tevatron run II. The finding of
this precision analysis are in qualitative agreement with previous results~\cite{Appelquist:2002wb}, but are at variance with the conclusions of a recent paper~\cite{Flacke:2005hb} where an analysis of LEP data including data from above the $Z$ pole and two loop electro-weak corrections to the $\Delta \rho$ parameter pointed to $R^{-1} \gtrsim 800$ (at 95\% C.L.).

Equally important turn out to be the lower bounds from the inclusive radiative decay $\bar{B}\to X_s\gamma$. It has been shown in ref.~\cite{Haisch:2007vb} that a refined analysis including in addition to the leading order contribution from the extra-dimensional KK states, the known next-to-next-to-leading order correction in the Standard model (SM) gives a lower bound on the compactification radius $R^{-1} \gtrsim 600$ GeV at $95\%$ confidence level (CL) and independent of the Higgs mass.

In this work we study the formation, production and possible detection  of bound states of Kaluza-Klein $n=1$ excitations  at $e^+e^-$ collisions. The production of bound states of KK excitation has been the object of some previous work~\cite{Carone:2003ms}.  As compared to ref.~\cite{Carone:2003ms} where the bound states production rates have been estimated by using a Breit-Wigner approximation, our study makes use of the method of the Green function in order to estimate the bound state contribution at the threshold cross-section, an effect
which can be as large as a factor of three when considering strongly interacting particles. In describing the interactions that allow the formation of level-1 KK bound states we assume that the  level-1 KK quark  excitations interact via an $\alpha_s$ driven Coulomb potential. This allows the  use of analytic expressions for  the Green function of the Coulomb problem but it should be kept in mind that the results and conclusions about formation and decay of the bound state depend on this assumption. This method has also been recently used by the present authors in a study of \emph{sleptonium} bound states within a slepton co-next to lightest supersymmetric particle (slepton co-NLSP) scenario of gauge mediated symmetry breaking (GMSB)~\cite{Fabiano:2005gt}.

The plan of the paper is as follows. Section~\ref{sec:UED} briefly describes
the UED model taken as a reference scenario. Section~\ref{sec:stateformation}
discusses the formation criteria and shows that the bound states of KK level-1
excitations do indeed form. Section~\ref{sec:greenfunction} describes the Green
function method for the bound states providing an analytic formula for the Born production cross section. The threshold cross section for the bound state is studied for several values of the scale of the extra dimension $R^{-1}$. Section~\ref{sec:decays} discusses
the possible decays of the bound states. Finally in Section~\ref{sec:detection}
we discuss the possible observation of the KK bound states at the $e^+e^-$
linear collider pointing to three possible signatures whose standard model background are also considered providing an estimate of the statistical significance. In section~\ref{sec:conclusions} we present the
conclusions.

\section{Universal Extra Dimensions}
\label{sec:UED}
The UED model is constructed considering the Standard Model
in a space time of $4+D$ dimensions, and assuming that all
SM particles are allowed to propagate in the extra
dimensions which typically are assumed to be compactified to a radius $R$.
In the following we follow strictly  the
notation of ref.~\cite{Hooper:2007qk}. We indicate the usual four dimensional
coordinates as $x^\mu$, $\mu=0,1,2,3$ and with $y^a, a=1, \cdots D$ the extra
space dimensions. The effective
four-dimensional Lagrangian is then obtained by dimensional reduction, i.e.
by integrating the $4+D$ dimensional SM Lagrangian over the $D$ extra space
dimensions. Thus one has:
\begin{widetext}
\begin{eqnarray}\label{xDlagrangian}
\nonumber {\mathcal L}_{eff}(x^\mu)&=&\int\ {\rm d}^Dy\Big\{-\sum_{i=1}^3\
\frac{1}{2\hat g^2_i}{\rm Tr}\Big[F_i^{AB}(x^\mu,y^a)F_{iAB}(x^\mu,y^a)\Big]+\\[0.2cm]
\nonumber &&+\left|(D_\mu+D_{3+a})H(x^\mu,y^a)\right|^2+\mu^2H^*(x^\mu,y^a)H(x^\mu,y^a)-
\lambda\big[H^*(x^\mu,y^a)H(x^\mu,y^a)\big]^2+\\[0.2cm]
\nonumber &&+i\left(\overline{ Q},\overline{ u},\overline{ d},\overline{ L},
\overline{ e}\right)(x^\mu,y^a)\left(\Gamma^\mu D_\mu+\Gamma^{3+a}D_{3+a}\right)
\left({ Q},{ u},{ d},{ L},{ e}\right)(x^\mu,y^a)+\\[0.2cm]
\nonumber &&\Big[ \overline{ Q}(x^\mu,y^a)\left(\hat\lambda_{ u}
\ { u}(x^\mu,y^a)i\sigma_2H^*(x^\mu,y^a)
+\hat\lambda_{ d}\ { d}(x^\mu,y^a)H(x^\mu,y^a)\right)+{\rm H.c.}\Big]+\\[0.2cm]
&&\Big[ \overline{ L}(x^\mu,y^a)\hat\lambda_{ e}\ { e}(x^\mu,y^a)H(x^\mu,y^a)+{\rm H.c.}\Big].
\end{eqnarray}
\end{widetext}
In the above  Eq.~\ref{xDlagrangian} $F_i^{AB}$ are the gauge field strength
tensors of SM gauge group $SU(2)\times U(1)\times SU(3)$ and $\hat{g}_i$ are
the gauge coupling constants in $(4+D)$-dimensions which have dimension of
$(mass)^{-D/2}$ as well as the Yukawa couplings $\hat{\lambda}_{u,d,e}$.
$D_\mu= \partial/\partial x^\mu -\mathcal{A}_\mu$ and
$ D_a= \partial/\partial y^a-\mathcal{A}_{3+a}$ are the covariant derivative
and $\mathcal{A}_{A} =-i\sum_{k=1}^3 \hat{g}_k T^r_i (\mathcal{A}_{A})_i^r$ are
gauge fields. $Q,L$ are the $SU(2)$ doublets, while $ u,d,e$  are the singlet
$(4+D)$-dimensional fermion fields. $\Gamma^A, A=0, \cdots (3+D)$ are
(4+D)-dimensional gamma matrices satisfying the anti-commutation relations
$\{\Gamma^A,\Gamma^B\}= 2g^{AB}$. In the following we shall deal with the
simplest case of only one extra-dimension ($D=1$)~\footnote{When D=1, one
can choose $\Gamma_\mu=\gamma_\mu$ and $\Gamma_4=i\gamma_5$}.

In order to extract the four-dimensional effective theory one needs to specify
how the extra-dimensions are compactified. It is found that in order to
reproduce chiral fermions in 4-dimension (the SM fermions) one is forced to
assume an orbifold compactification structure which depends on the number of
extra dimensions. For D-odd (e.g. $D=1$) one chooses an $S^1/Z_2$ or\-bi\-fold
structure with $Z_2$ being the reflection symmetry $y \to -y$.
One assumes that the gauge fields ${\mathcal A}_\mu$ and the Higgs boson $H$
are even under the $y\to-y$ transformation, while the ${\mathcal A}_5$ is
assumed odd. This results in a Fourier series expansion of the fields which
defines the zero modes (that correspond to the SM particles) and the level-$n$
KK-excitation (coefficients of the expansions).

\begin{eqnarray}
H(x^\mu,y)&=&\frac{1}{\sqrt{\pi R}}\Big[H^0(x_\mu)+\sqrt{2}
\sum_{n=1}^\infty H^n(x_\mu)\cos\left(\frac{ny}{R}\right)\Big]\cr
{\mathcal A}_\mu(x^\mu,y)&=&\frac{1}{\sqrt{\pi R}}
\Big[{\mathcal A}_{\mu}^0(x_\mu)+\sqrt{2}\sum_{n=1}^\infty
{\mathcal A}_{\mu}^n(x_\mu)\cos\left(\frac{ny}{R}\right)\Big]\cr
{\mathcal A}_5&=&\sqrt{\frac{2}{\pi R}}\sum_{n=1}^\infty
{\mathcal A}_5^{(n)}(x_\mu)\sin\left(\frac{ny}{R}\right).
\end{eqnarray}
As already anticipated, difficulties arise when trying to construct chiral
fermion fields in more than four dimensions. This is because  in
$5$-dimensions, for example, it is not possible to construct the equivalent of
the $\gamma_5$ matrix and  bi-linear quantities like $\bar{\psi}\gamma^\mu\gamma_5\psi$
are not invariant under $5-$dimensional Lorentz transformations and therefore they cannot
appear in the $5D$-Lagrangian. This ultimately implies that for each standard model field
one must introduce two $5D$ fermion fields whose zeroth order modes combine to give the
$4D$ chiral fermion. This however leaves some extra massless degrees of freedom at the
zero level which can only be eliminated by formulating the theory on an
orbifold~\cite{Macesanu:2005jx,Papavassiliou:2001be}. The 5-dimensional fermion field is thus expanded as:
\begin{eqnarray}
\Psi (x^\mu,y)&=&\frac{1}{\sqrt{\pi R}}\Big[\psi^{\rm SM}(x^\mu)+\sqrt{2}\sum_{n=1}^\infty \psi_{L}^{(n)}(x^\mu)\cos\left(\frac{ny}{R}\right)\cr
&&\phantom{xxxxxxxxxxx}+\psi_{R}^{(n)}(x^\mu)\sin\left(\frac{ny}{R}\right)\Big].
\end{eqnarray}

Performing the integration over the extra space dimension the derivatives with
respect to $y$ will bring about mass terms that scale with the compactification
radius $R$. Every level-$n$ KK excitation $X_{(n)}$ acquires, in addition to
 the SM mass (level $0$), obtained via the Higgs mechanism, a new term:
\begin{equation}
m^2_{X^{(n)}} =  m^2_{X^{(0)}}+\frac{n^2}{R^2} ,
\label{tree}
\end{equation}

These relations are however modified by radiative corrections which turn out to
 be cut-off dependent. These radiative corrections arise from loop diagrams
 traversing the extra-dimension~\cite{Cheng:2002ab} (bulk loops) and from
 kinetic terms localized on the brane which appear on the or\-bi\-fold structure.
\begin{eqnarray}\label{radcorrections}
\delta(m^2_{B^{(n)}}) &=& \frac{g'^2}{16 \pi^2 R^2}
  \left[ \frac{-39}{2} \frac{\zeta(3)}{\pi^2} -\frac{n^2}{3} \log \,
  (\Lambda R) \right] \nonumber \\
\delta(m^2_{W^{(n)}}) &=& \frac{g^2}{16 \pi^2 R^2} \left [
  -\frac{5}{2} \frac{\zeta(3)}{\pi^2} + 15 n^2 \log \,
  (\Lambda R) \right ]\nonumber \\
\delta(m^2_{g^{(n)}}) &=& \frac{ g_3^2}{16 \pi^2 R^2} \left [
  -\frac{3}{2} \frac{\zeta(3)}{\pi^2} + 23 n^2 \log \,
  (\Lambda R) \right ]\nonumber \\
\delta(m_{Q^{(n)}}) &=& \frac{n}{16 \pi^2 R} \left [ 6 g_3^2+ \frac{27}{8}
g^2 + \frac{1}{8} g'^2 \right] \log \, (\Lambda R) \nonumber \\
\delta(m_{u^{(n)}}) &=& \frac{n}{16 \pi^2 R} \left [ 6 g_3^2+
 2 g'^2 \right] \log \, (\Lambda R \nonumber) \\
\delta(m_{d^{(n)}}) &=& \frac{n}{16 \pi^2 R} \left [ 6 g_3^2+  \frac{1}{2}
g'^2 \right] \log \, (\Lambda R \nonumber) \\
\delta(m_{L^{(n)}}) &=& \frac{n}{16 \pi^2 R} \left [ \frac{27}{8}
g^2 + \frac{9}{8} g'^2 \right] \log \, (\Lambda R) \nonumber \\
\delta(m_{e^{(n)}}) &=& \frac{n}{16 \pi^2 R}\frac{9}{2} g'^2 \log \, (\Lambda R) \; .
\end{eqnarray}

Here $\zeta(z)$ is the Riemann zeta function, $\zeta(3)\approx 1.2020$, and
$\Lambda$ is the cutoff scale of the theory. It correspond to the energy scale
at which  the effective 5-dimensional theory will break down, that is where the
5-dimensional couplings become strong and the theory is no longer perturbative.
$\Lambda$ is the only additional parameter of the UED model beside the size of
the extra dimension $R$. It can be estimated requiring that that loop expansion
parameters remain perturbative. It has been found that the SU(3) interaction
becomes non perturbative before the other gauge interactions for values of
$\Lambda\gtrsim 10\, R^{-1}$.  So the particle spectrum is typically computed
with the above Eq.~\ref{radcorrections} taking $\Lambda R = 5,10,20$. In the
above expressions the brane kinetic terms are those dependent on the cutoff
scale $\Lambda $. It should be noted that for KK scalars and spin-1 bosons the
corrections in Eq.~\ref{radcorrections} simply add to Eq.~\ref{tree}, while the
corrections for the fermion masses are introduced via the replacement $n^2/R^2
\to (n/R +\delta m^{(n)})^2$.

\begin{table*}[t!]
\begin{ruledtabular}
\begin{tabular}{cccccc}
$R^{-1}$ (GeV)&$KK$ {mass} (GeV) & $\alpha_s(r_B^{-1})$& State mass $M$ (GeV)& $E_{1S}$ (GeV) & $\Delta E(2P-1S)$ (GeV) \\
\hline
300 & 358.54 &0.136&714.06&  2.937  & 2.203 \\
400 & 478.05 &0.131&952.57&   3.627  & 2.720 \\
500 & 597.56 &0.127&1190.92&   4.279  & 3.209 \\
600 & 717.08 &0.124&1429.30&  4.903  & 3.677 \\
700 & 836.60 &0.122&1667.69&  5.505  & 4.128 \\
800 & 956.11 &0.120&1906.11&  6.089  & 4.567 \\
900 & 1075.62 &0.118&2144.54& 6.658  & 4.993 \\
1000& 1195.14 &0.116&2382.98& 7.214  & 5.411 \\
\end{tabular}
\end{ruledtabular}
\caption{\label{table1}Results of Coulombic model for the bound state of the level-1 iso-doublet $U_1$ quark. The strong coupling $\alpha_s $ is computed at the scale $Q=r_B^{-1}$, where $r_B=3/(2m\alpha_s)$ is the Bohr's radius. For each mass value  $m$ the scale $Q=r_B^{-1}$ depending itself on $\alpha_s$ must be solved numerically from the equation $Q=(2/3)\, m \,\alpha_s(Q)$.}
\end{table*} 
\section{ $\mathbf{u_1 \overline{u}_1}$ Bound State Formation}
\label{sec:stateformation}

In this section we shall review the possible creation of a
bound state of the level-1  $KK$-excitation of the
$u$-quark, i.e. a bound state  $u_1 \bar{u}_1$. The
interaction among two Kaluza--Klein excitations are driven
by the QCD interaction, thus bearing no differences with
respect to the Standard Model; the strength of the
interaction is given by $\as$ computed at a suitable
scale~\cite{Fabiano:1993vx,Fabiano:1994cz,Fabiano:1997xh}.
We shall adopt the same formation criterion stated there,
namely that the formation occurs only if the level
splitting depending upon the relevant interaction existing
among constituent particles is larger than the natural
width of the would--be bound state. This translates into
the formation requirement \beq \Delta E_{2P-1S} \ge \Gamma
\label{eq:formation} \eeq where $\Delta E_{2P-1S} =
E_{2P}-E_{1S}$ and $\Gamma$ is the width of the would--be
bound state. The latter is twice the width of the single
$KK$ quark, $\Gamma = 2 \Gamma_{KK}$, as each $KK$ quark
could decay in a manner independent from the other.

We shall stress that $\Gamma$ bears no resembling to the total decay width of
the resonance, as it only includes the single $KK$ decay mode and not the
annihilation modes discussed later in sec.~\ref{sec:decays}. It represents the
minimal energy level spread needed for bound state formation, which allows for
the separation among the fundamental and the first excited state. If, and only
if, the bound state is formed, then it is possible to discuss its annihilation
widths as described in sec.~\ref{sec:decays}.
In our model $V(\bm{x})$ is given by  a Coulombic potential
\beq
V(r) = - \frac{4 \as}{3r}
\label{eq:coulomb}
\eeq
with $r=|\bm{x}|$, and
where $\as$ is the usual QCD coupling constant which
has been taken at a suitable scale as described
in~\cite{Fabiano:1993vx,Fabiano:1994cz}. This model has proved to be
reliable because of high mass values involved in the problem, and gives the
great advantage of having full analytical results.
We are thus able to compute its energy levels given by the expression
\beq
\varepsilon_n = - \frac{4 }{9} \frac{m\as^2}{n^2}
\label{eq:ebound}
\eeq
and the separation of the first two energy levels is given by
\beq
\Delta E_{2P-1S} = \frac{1}{3} m \as^2
\label{eq:deltae}
\eeq
The scale at which $\as$ is evaluated is given by the inverse of
Bohr's radius $r_B=3/(2m\as)$, the average distance of the constituents of the bound state. Therefore it is found by solving numerically the equation $Q=(2/3) \, m \, \as (Q)$ and it is of
order $\mathcal{O}(10\mbox{ GeV})$ for $m\approx 300$ GeV and $\mathcal{O}(100\mbox{ GeV})$ fr $m\approx 1200$ (GeV). The corresponding values of $\as$ are given in Table~\ref{table1}. The mass of the $n$th bound state is
given by the expression:
\beq
M_n = 2m+\varepsilon_n
\label{eq:massbound}
\eeq
where $m$ is the mass of the constituent $u_1$ quark and $E_n$ is given
by~(\ref{eq:ebound}). The wavefunction at the origin, which will be needed
in order to compute decay widths, for this particular model is given
by the expression
\beq
|\psi(0)|^2 = \frac{1}{\pi}\left ( \frac{2}{3}m\as\right )^3
\label{eq:psizero}
\eeq
The obtained results are given in Table~\ref{table1}.

We observe that the bound state energies are of the order of the GeV for this
range of $KK$ mass, and that the spreading of the first two bound states raise
linearly with $m$.

In order to determine whether the bound state will be formed we shall apply the
criterion given in eq.(~\ref{eq:formation}). The $KK$-quark decay widths have
been already computed in~\cite{Carone:2003ms}, where it has been shown that
their values are at most of the order of 100~MeV, one order of magnitude less
than the energy splittings. In this scenario the eq.~(\ref{eq:formation})
requirement is always fulfilled, and the  bound state is formed for $KK$-quark
masses in this investigation range.

\section{Green Function}
\label{sec:greenfunction}

In order to describe the cross--section of a $KK$ bound
state in the threshold region we shall use the method of
the Green function. We briefly review here the essential
features of the mechanism, and refer the reader to the
literature for further details we~\cite{Fabiano:2001cw}. We
start from the Schr\"odinger equation which describes the
bound state by means of a suitable potential $V(\bm{x})$,
\beq
\bm{H}\psi= \left (- \frac{\nabla_{\bm{x}}^2}{2m} +V(\bm{x}) \right
) \psi=\mathcal{E}\psi \; ,
\label{eq:schroedinger}
\eeq
where $\mathcal{E}$ is the energy eigenvalue of the bound state.
The threshold cross--section of this bound state is then
proportional to the imaginary part of the $S$ wave Green
function of this Schr\"odinger equation,
$\mathcal{G}_{1S}(\bm{x},\bm{y},E)$,  where the two
constituent particles are sited in $\bm{x},\bm{y}$ and $E$
is the energy offset from the threshold (not to be confused
with $\mathcal{E}$),
\beq
(\bm{H}-E) \mathcal{G}_{1S}(\bm{x},\bm{y},E) = \delta(\bm{x}-\bm{y})
\label{eq:defgreen}
\eeq

By means of the
substitution  $E \to E + i \Gamma$ we take into account the
finite width of the state.

The cross--section is thus proportional to the expression
\beq
\sigma \sim  Im \left . \left [ Tr \frac{\partial}{\partial x_i}
\frac{\partial}{\partial y_j}\mathcal{G}_{1S}(\bm{x},\bm{y},E)
\right ] \right |_{\bm{x}=0,\bm{y}=0}
\label{eq:tracegreen}
\eeq
the derivative of eq.~(\ref{eq:tracegreen}) has a simple expression, as we
have
\beq
{\sf Tr}\, \left .\frac{\partial}{\partial {x}_i}\, \frac{\partial}{\partial
{y}_j}\, \mathcal{G}_{1S}(\bm{x},\bm{y},E) \right |_{\bm{x}=0,\bm{y}=0} =
9\mathcal{G}_{1S}(0,0,E)\,\, .
\label{eq:g1der}
\eeq

\begin{figure*}[t!]
\includegraphics*[scale=1.05]{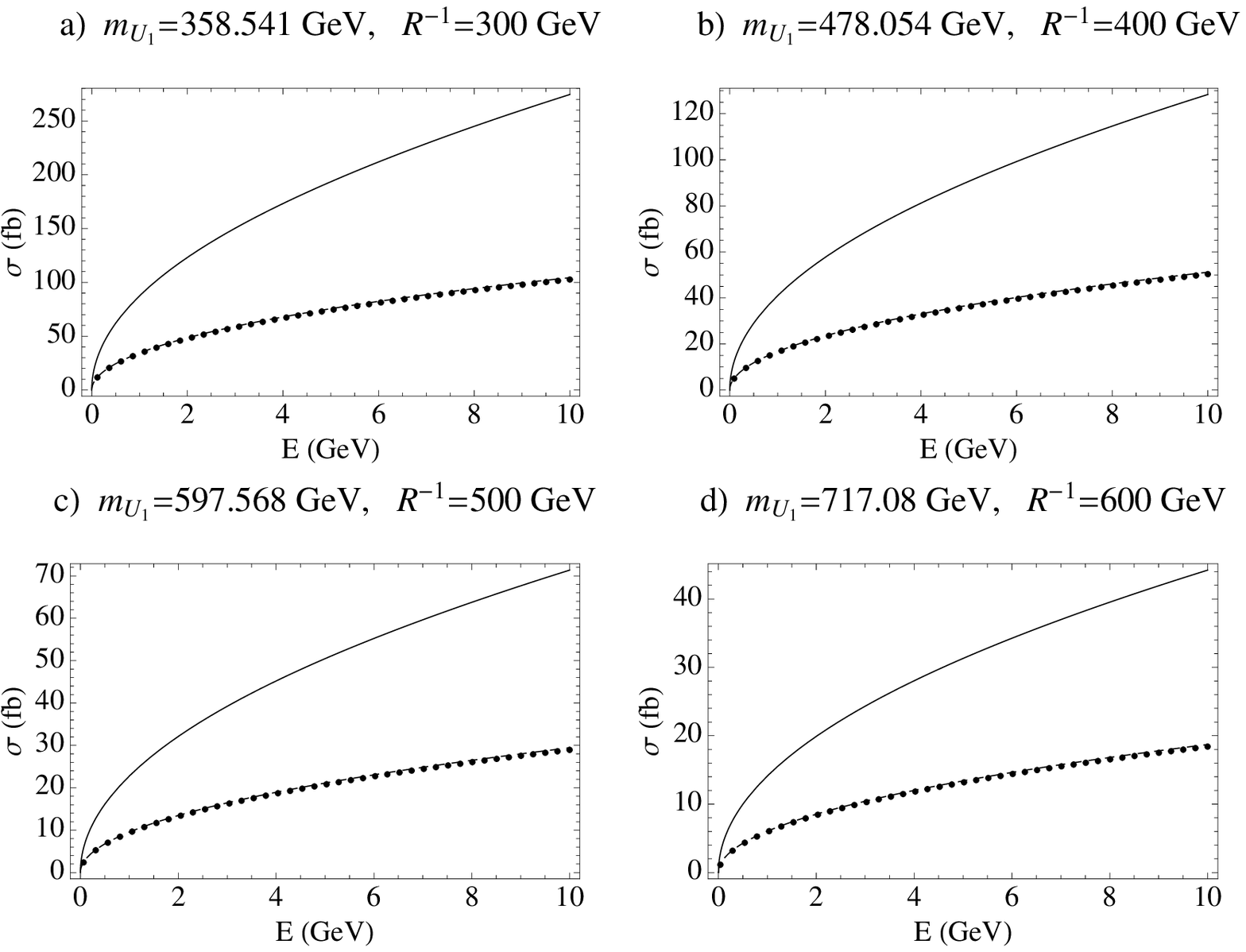}
\caption{\label{fig1}Production cross sections  of level-1
KK doublet quark  bound states $U_1 \bar{U}_1$ as a function of the energy offset from threshold ($\sqrt{s} = 2m_{U_1} +E$), for values of the  scale of the extra-dimension $R^{-1}$ in the range $[300\div600]$~GeV and a total width of $\Gamma=0.5$~GeV. The continuous line is the Green Function result, the dotted one is the Born approximation given by our analytical formula (Eq.~\ref{eq:sigmaborn}). The full circles represent the Born cross section from  the CalcHEP~\cite{Datta:2010us} numerical session inlcuding also the annihilation diagrams of $\gamma_2$ and $Z_2$ whose contribution is however completely negligible. The numerical results from CalcHEP are in complete agreement with our analytical formula in Eq.~\ref{eq:sigmaborn}. The cut-off scale $\Lambda$, at which perturbative expansions break down, has been fixed so that $\Lambda R =20$.}
\end{figure*}
\begin{figure*}[t!]
\includegraphics[scale=1.05]{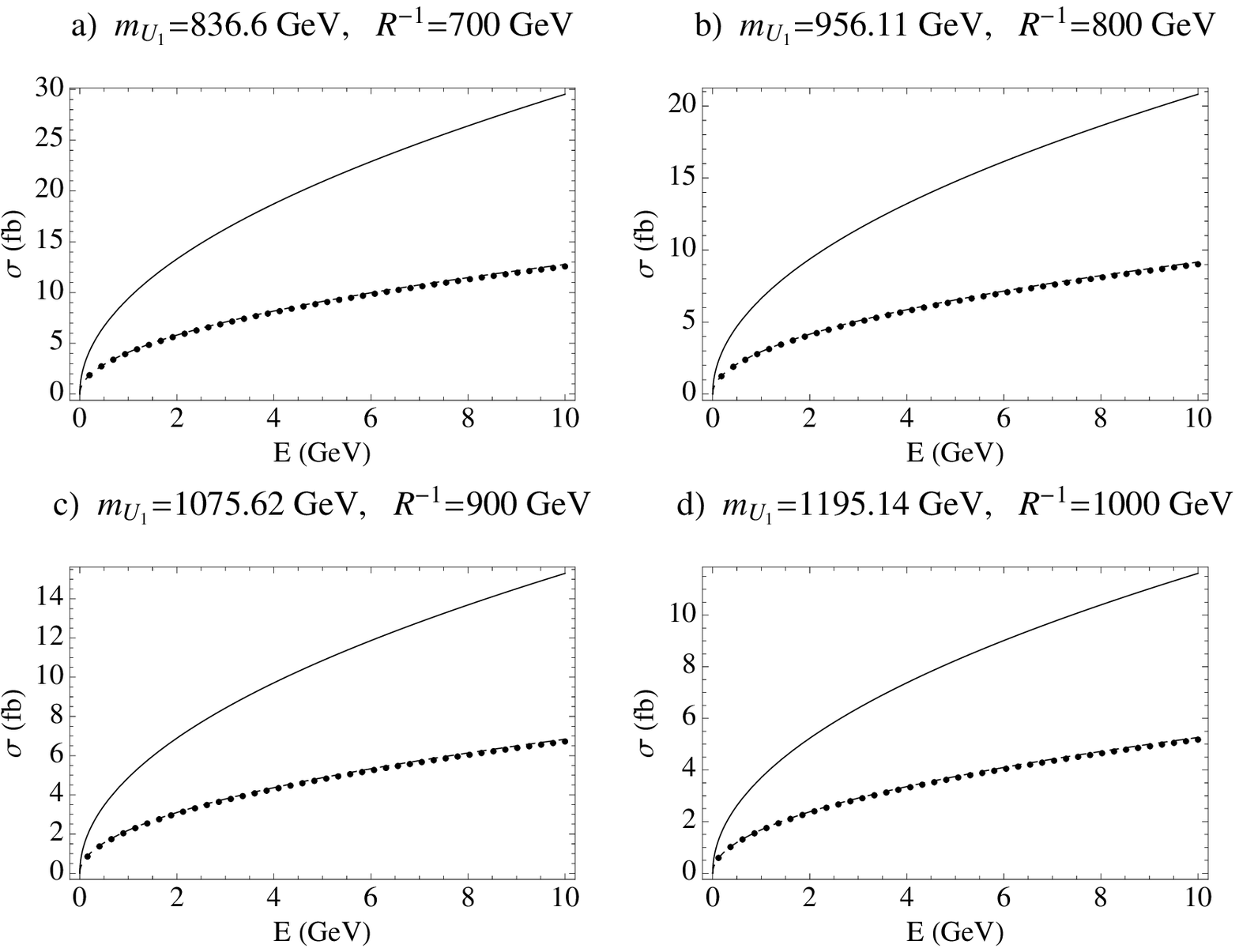}
\caption{\label{fig1new}Production cross sections  of level-1
KK doublet quark  bound states $U_1 \bar{U}_1$ as a function of the energy offset from threshold ($\sqrt{s} = 2m_{U_1} +E$), for values of the  scale of the extra-dimension $R^{-1}$ in the range $[700\div1000]$~GeV and a total width of $\Gamma=0.5$~GeV. The continuous line is the Green Function result, the dotted one is the Born approximation given by our analytical formula (Eq.~\ref{eq:sigmaborn}). The full circles represent the Born Cross section from the CalcHEP~\cite{Datta:2010us} numerical session inlcuding also the annihilation diagrams of $\gamma_2$ and $Z_2$ whose contribution is however completely negligible. The numerical results from CalcHEP are in complete agreement with our analytical formula in Eq.~\ref{eq:sigmaborn}. As in Fig.~\ref{fig1} we have kept fixed $\Lambda R = 20$.}
\end{figure*}

The complete expression for the $1S$ Green function of our problem
as a function of energy from threshold
is given with a slight change of notation by~\cite{Penin:1998mx}:
\beqy
\mathcal{G}_{1S}(0,0,E+i\Gamma) &=& \frac{m}{4 \pi}
\left [-2 \lambda \left ( \frac{k}{2 \lambda}  +
\log \left( \frac{k}{\mu} \right ) + \right . \right . \nonumber \\
 &&\left .  \left . \phantom{xxx\frac{1}{2}}\, \psi(1-\nu) +2 \gamma-1  \right ) \right ]
\label{eq:green1s} \eeqy where $ k = \sqrt{-m (E+i
\Gamma)}$, $ \lambda = 2 \as m/3$ and the wave number is $ \nu
=\lambda/k $; $E=\sqrt{s}-2m$. The $\psi$ is the
logarithmic derivative of Euler's Gamma function
$\Gamma(x)$, $\gamma \simeq 0.57721$ is Euler's constant
and $\mu$ is an auxiliary parameter coming out from a
dimensional regularization, the factorization scale, that
cancels out in the determination of physical observables.

The final expression for the production cross--section of a $KK$ bound state is thus given by
\beq
\sigma(m,E,\Gamma,\as) = \frac{18 \pi}{m^2} \sigma_B\,\text{Im} \, \left[\mathcal{G}_{1S}\right]
\label{eq:sigmagreen}
\eeq
where $\sigma_B$ is the Born expression of the
cross--section~\cite{Burnell:2005hm}
The process $e^+e^- \to U_1 \bar{U}_1$ proceeds through the annihilation into the standard model (level-0) gauge bosons $\gamma $ and $Z$ but in principle one should also consider the contribution of the  level-2 gauge bosons $\gamma_{(2)}$ and  
$Z_{(2)}$.  Especially so in our case of threshold production of the pair $u_1 \bar{u}_1$. Indeed in this case $m\approx 1/R$, and $\sqrt{s} =2m +E \approx 2/R +E$ and since  $m_{\gamma_2}\approx 2/R$ when producing at threshold the $u_1\bar{u}_1$ pair we would be close  to the $\gamma_2$ and $Z_2$ resonances.  However as discussed in section~\ref{sec:UED} the mass spectrum is modified by the radiative corrections.  We have verified that over the region of parameter space $ 300\,\,  \text{GeV }\leq R^{-1} \leq 1000\,\,  \text{GeV}$ and  $2 \leq \Lambda R \leq 70$
the pair production threshold $2m_{u_1}$ is always larger than $m_{\gamma_2}, m_{Z_2}$ and thus these resonances should in principle be included in the calculation.  We have also verified, cross checking our calculation with the output of a CalcHEP~\cite{Pukhov:2004ca,Datta:2010us} session, that the numerical impact of these diagrams is  completely negligible. Their contribution turns out to be five orders of magnitude smaller than that of the SM gauge bosons $\gamma,Z$. The analytic formula of the Born pair production cross section $e^+e^- \to \gamma^*,Z^* \to U_1\bar{U}_1$ can be deduced for example from those  of heavy quark ($t \bar{t}$) \cite{Panella:1993zk} taking into account the fact the the level-1 KK quarks are \emph{vector-like} i.e. their coupling to the $Z$ is of the $\gamma^\mu$ type and has no axial component. Following the notation of \cite{Panella:1993zk} the amplitude is written as:
\begin{eqnarray*}
{\cal M}&=& g_V^2\sum_{V=\gamma , Z}\bar{v}(k_2) \gamma^\mu(a_V+b_V\gamma^5)u(k_1)\times  \\ & &\frac{1}{D_V(s,M_V)}\, \bar{u}(p_1) \gamma_\mu (A_V+B_V\gamma^5)v(p_2) 
\end{eqnarray*}
where $D_V(s)= s-M_V^2 +i M_Z \Gamma_Z$ is the gauge boson propagator factor, $a_V , b_V$ are the (\emph{standard model}) coupling coefficients of the electron to the gauge bosons while $A_V, B_V$ are the coupling coefficients of the level-1 KK  $U_1$ quark to the gauge bosons. These electron coefficients are: $a_\gamma =-1$, $b_\gamma=0$, $a_Z=-1/4+\sin^2\theta_W$, $b_Z=+1/4$, with $\theta_W$ the Weinberg angle of the $SU(2)\otimes U(1)$ gauge theory. The $U_1$ coefficients are: $A_\gamma= +2/3$, $A_Z= 1/2 -(2/3)\sin^2\theta_W$~\cite{Hooper:2007qk}, $B_\gamma=B_Z=0$ (recall that $U_1$ is vector-like). Finally $g_\gamma=e$ and $g_Z=e/(\cos\theta_W\sin\theta_W)$ with $e$ the electronic charge. The final expression of the Born pair production cross section is:   
\begin{eqnarray}
\label{eq:sigmaborn}
\sigma_B&=&\sigma_\gamma +\sigma_Z +\sigma_{{\gamma}Z}\\ 
\sigma_\gamma&=&\frac{16\pi \alpha^2}{9s^2}\beta (s+2m^2)\\
\sigma_{Z}&=& {4\pi\alpha^2}\,\frac{(a_Z^2 +b_Z^2) A_Z^2}{[c_w^2s_w^2]^2}\,\frac{\beta (s+2m^2)}{|D_Z(s)|^2}\\
\sigma_{\gamma Z}&=& -8\pi\alpha^2 \frac{a_ZA_\gamma A_Z}{c_w^2s_w^2}\,\frac{\beta (s+2m^2)}{|D_Z(s)|^2} (1-\frac{M_Z^2}{s})
\end{eqnarray}
where $c_W=\cos\theta_W$ and $s_W=\sin\theta_W$ and $\alpha=e^2/(4\pi)$ is the QED fine structure constant.
From Eq.~(\ref{eq:green1s}) one can readily see the behavior of the cross--section~(\ref{eq:sigmagreen}) for large $E$  is given by $k$. The finite width of the state has been taken into account by the substitution $E \to E + i \Gamma$, and
this position makes a great quantitative difference below threshold. 
When computed for positive energy offset the variation of $\Gamma$ 
makes essentially no difference for the resulting cross--section.

In this work we shall concentrate on the continuum region of the
cross--section, namely $E>0$. The region below threshold, $E<0$, has been
already discussed in detail in ref.~\cite{Carone:2003ms}, where the authors presented an analysis of both the  positions and the widths of the peaks using a Breit-Wigner description.
In this respect the Green function approach  does not carry substantial
differences relative to the Breit--Wigner one. Indeed the
position of the poles  and the broadening  of the peaks
are the same due to the presence, inside the $\psi$ function of
Eq.~(\ref{eq:green1s}), of terms which include the binding energy $E_n$
of Eq.~(\ref{eq:ebound}) and the decay width $\Gamma$.
For values of $E$ close to $E_n$, the argument of the $\psi$ function inside
Eq.~(\ref{eq:green1s}), namely $(1-\nu)$, approaches a negative integer,
simple pole of the function in the complex plane, while the presence
of the $\Gamma$ determines the width of the peak centered in $E_n$.

In Fig.~\ref{fig1}  we  show the cross--section
for a range of the value of the scale of the extra dimension, $R^{-1}=300-600$~GeV, while Fig.~\ref{fig1new} provides the same plots are shown the range
$R^{-1}=700-1000$~GeV. In both figures the value of the other parameter is fixed at $\Lambda R =20$. This parameter enters our calculations only when computing the mass spectrum through the logarithmic terms in Eq.~\ref{radcorrections}. We thus provide a quantitative study of the effect of the formation of bound states of the level-1 KK quarks with respect to the parameter of the model ($R^{-1}$). The results are less sensitive to the other parameter ($\Lambda R$) which only enters through the logarithmic factors in the radiative correction terms in the mass spectrum of the model. In Figs.~\ref{fig1}\&~\ref{fig1new} we have fixed $\Lambda R=20$ and varied $R^{-1}$ computing the corresponding values of the level-1 KK quark mass, and assuming the energy of the collider being fixed at $\sqrt{s}=2m_{U_1}+ E$, $E$ being the energy offset from the threshold. 
We have used a value of
$\Gamma = 0.5$~GeV for illustrative purpose, compatible with the formation of
 bound state. Different choices of $\Gamma$ by even two orders of magnitude
smaller will not make a visible difference on the figures.

One can observe that the cross--section obtained by the 
Green function has a behavior like $\sqrt{E}$ for small
energy offset. The cross--section decreases with increasing
mass: for $E=10$~GeV its value is about 250~fb at
$R^{-1}=300$~GeV, goes down to approximately 70~fb at
$R^{-1}=500$~GeV. Finally it approaches 13~fb at $R^{-1}=1000$ GeV as
can be seen form Fig.~\ref{fig1new}. The Green function cross section  
is larger than the Born cross-section by a factor that 
ranges from 2.7  (at $R^{-1}=300$ GeV) down to 2.2 (at $R^{-1}=1000$ GeV)  
at the same energy offset value ($E=10$ GeV). This result is due to the fact that
the Green function method takes into account the existing
interaction among constituent particles, and the
contribution of binding energies accumulate towards the
$\varepsilon =0$ level, thus substantially contributing to the
continuum region as well.

An important consideration is in order here. The results for the bound state
cross--section given depend solely on the coupling constant $\as(r_B)$ and
the mass of the $KK$ excitation, thus they are universal to bound states
made out of other flavors of $KK$ quarks. This however does not apply to other
kinds of $KK$ excitations bound states, like for instance bound states of
KK-leptons. In this case we have $\alpha\equiv\alpha_{QED}$ for coupling
constant, much weaker at this scale than the strong coupling constant
$\as(r_B)$. The QED coupling constant  would lead, not only to lower values for
the absolute production cross-section, but will also reduce drastically the main
effect being discussed here, i.e. the enhancement, above threshold, due to the
bound state interaction relative to the Born cross-section. The threshold
cross-section  would be only a few percent larger than the Born cross-section.
It would not clearly be an effect as large as that  shown in
figs~(\ref{fig1}),~(\ref{fig1new}) which turns out to be quite striking, i.e.
the threshold bound-state cross section is about three times as large as the
Born result.

\section{$\mathbf{u_1 \overline{u}_1}$ Decay Widths}
\label{sec:decays}
\begin{figure}[t!]
\includegraphics*[scale=0.8]{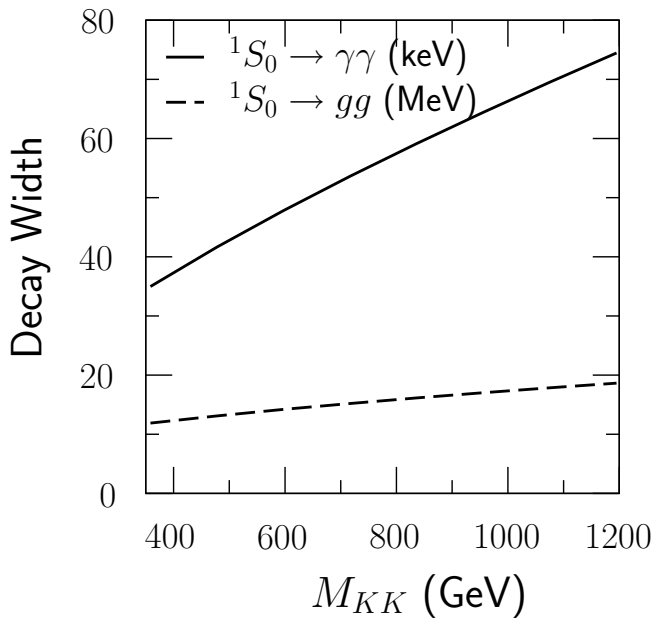}
\caption{\label{fig2}\emph{Solid line}: decay width (keV)
of the pseudoscalar $u_1 \bar{u}_1$ bound state to two
photons as a function of $KK$ excitation mass. \emph{Dashed
line}: decay width (MeV) of the pseudoscalar $u_1
\bar{u}_1$ bound state to two gluons as a function of the
$KK$ mass. 
}
\end{figure}
\begin{figure}[t!]
\includegraphics*[scale=0.8]{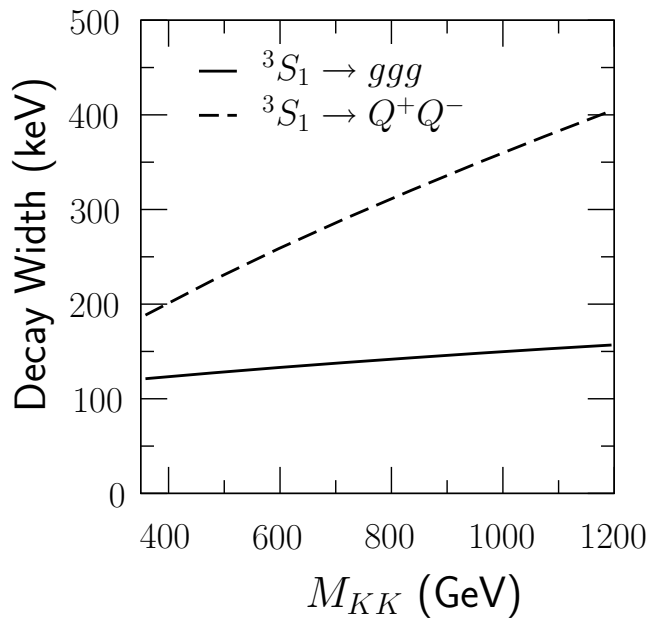}
\caption{\label{fig3} Decay widths of the $u_1 \bar{u}_1$
vector bound state to two charged particles and into three
gluons as a function of $KK$ mass. Here we have considered
all possible e.m. decay channels. }
\end{figure}
The $KK$ bound states we discuss here are the pseudoscalar $^1S_0$ and
the vector one $^3S_1$. For the pseudoscalar state the decay channels are
into two photons or two gluons for which the following Born level expressions
hold (see for instance~\cite{Fabiano:1994cz}):
\beq
\Gamma_B(^1S_0 \to \gamma \gamma) =   q_i^4\alpha^2 \frac{48 \pi |\psi(0)|^2}{M^2}
\label{eq:scalarto2photons}
\eeq
and
\beq \Gamma_B(^1S_0 \to gg) =   \as^2 \frac{32 \pi
|\psi(0)|^2}{3M^2} \; . \label{eq:scalarto2g}
\eeq
Here $q_i$ is the charge of the constituent quark of the bound state,
while $M$ and $|\psi(0)|^2$ are given by~(\ref{eq:massbound})
and~(\ref{eq:psizero}) respectively.

The QCD
radiative correction~\cite{Kwong:1987ak}, which is the same
in the two cases, lead to the following one-loop width:
\beq \Gamma = \Gamma_B \left [ 1 + \frac{\as}{\pi} \left (
\frac{\pi^2-20}{3}  \right ) \right ]
\label{eq:pseudoscalar1loop} \eeq The results obtained for
the two decays $^1S_0 \to \gamma \gamma$ and $^1S_0 \to gg$
are shown in Fig.~\ref{fig2}.

For the vector case $^3S_1$ the relevant decay channels are the one in charged
pairs and the one into three gluons, for which one has
\beq
\Gamma_B(^3S_1 \to q_f^+q_f^-) =   q_i^2 q_f^2 \alpha^2 \frac{16 \pi |\psi(0)|^2}{M^2}
\label{eq:vectortoee}
\eeq
and
\beq
\Gamma_B(^3S_1 \to ggg) = \frac{(\pi^2-9)}{\pi}  \as^3 \frac{160 \pi |\psi(0)|^2}{81M^2}
\label{eq:vectorto3g}
\eeq
The charge of the final state charged particle is given by $q_f$.
The QCD radiative corrections~\cite{Kwong:1987ak} modify these expressions into
\beq
\Gamma(^3S_1 \to q_f^+q_f^-) =
\Gamma_B(^3S_1 \to q_f^+q_f^-) \left ( 1- \frac{16}{3} \frac{\as}{\pi} \right )
\label{eq:vectortoee1loop}
\eeq
and
\begin{eqnarray}
\Gamma(^3S_1 &\to& ggg) = \Gamma_B(^3S_1 \to ggg)  \times  \nonumber \\
&&\times \left\{ 1+   \displaystyle\frac{\as}{\pi}   \left [ -14
+\frac{27}{2} \left ( 1.161+ \log(2) \right ) \right ]  \right \} \nonumber \\
\label{eq:vectorto3g1loop}
\end{eqnarray}
Observe that the $\as$ that appears in the perturbative corrections has to
be computed at a scale of the order of $2m$. It is thus different from
the $\as$ occurring in the expression of the wavefunction at the origin given
by~(\ref{eq:psizero}), the latter being computed at the scale of the inverse
of Bohr radius.

The two decays of the vector state are shown together in
Fig.~\ref{fig3}.

We observe that only the pseudoscalar hadronic decay is in the MeV range
and raises approximately linearly with $KK$ mass. The $^1S_0$ photonic decay
and $^3S_1$ decays are smaller by almost two orders of magnitude for the
considered $KK$ mass range.
For the pseudoscalar case the hadronic is the dominant decay by far, while
in the vector case the decay into charged particles, when taking into account
all possible processes as seen in Fig.~\ref{fig3} overtakes the
hadronic decays.

Other electro-weak decay channels are negligible. Those are proportional to
$\alpha^2$, thus their ratio to gluonic decays is suppressed by
$(\alpha/\as)^2$, at least by two orders of magnitude.

For most scenarios depending upon the values of $\Lambda$ and
$R$~\cite{Carone:2003ms} single quark decay becomes the dominant decay channel
for the bound state.

Any two--body decay width of the bound state is proportional
to $g^2|\psi(0)|^2/M^2$, where $g$ is the relevant coupling to the decay
particles. Thus any electro-weak width is in the keV range, as previously
seen. This result is true in general for any two--body decay process,
the only notable exception being the hadronic decay of~(\ref{eq:scalarto2g}),
as the $g$ coupling this time is rather large, being equal to $\as$. In this
case the value is in the MeV range, as seen in Fig.~\ref{fig2}.

Three--body decays are further suppressed with respect to previous formula
by another power in $g$ and phase--space reduction, resorting again in the
keV range of energies.

From~\cite{Carone:2003ms} one sees that in most cases single quark decays (SQD)
are by far the most important decay channels of the bound state, to the order
of hundreds of MeV, while as discussed above bound state decays are
essentially negligible.
Moreover a comparison of those SQD widths with the results of
Table~\ref{table1} through eq.~(\ref{eq:formation}) shows that
for the considered mass range of $KK$ there is formation of the
bound state.
\section{Detection}
\label{sec:detection}
As we have previously seen, for large $R^{-1}$ values ($R^{-1}> 300$~GeV) SQD is the dominant decay channel for a $KK$ bound state, thus leading a to a dominant signature consisting of two monochromatic quarks plus missing energy. 
Other interesting signatures that could be considered, for example the three jet production due to the bound state decay into three gluons  $^3S_1 \to ggg $ discussed in the previous section, are clearly subdominant given the fact that ${\cal B}(^3S_1 \to ggg ) \approx 10^{-3}$~\footnote{The value ${\cal B}(^3S_1 \to ggg ) \approx 10^{-3}$ is easily obtained assuming a SQD width of the order of a few hundreds MeV and combining this with the results of the partial widths given for example in Fig.~\ref{fig3} (or by using Eq.~\ref{eq:vectortoee} and Eq.~\ref{eq:vectorto3g}).}.  Such signatures, in addition would  have to be confronted with important QCD backgrounds. Therefore in the following discussion we concentrate on the dominant channels given by the single quark decay whose branching fractions, on the contrary, can be as high as 65\% and 98\% and involve missing energy  in the final state.
Following~\cite{Fabiano:2001cw} we limit our analysis to the region above
threshold,  \emph{i.e.} $E>0$.
The region below threshold, $E<0$,  is characterized by peaks in the cross
section for values of $E$ equal to binding energies of the bound states.
The width of those peaks are given by the decay width of the bound state, which are
at most of the order of the MeV for the SQD and much less, of the order of the
keV, for other annihilation decay modes, as discussed
in sect.~\ref{sec:decays}.

From Eq.~(\ref{eq:ebound}) we can estimate the separation
of the various peaks below threshold, which tend to merge
when they accumulate, that is for $n$ such that \beq
\frac{4}{9} m  \as^2\left [\frac{1}{n^2} -
\frac{1}{(n+1)^2} \right ] \sim \Gamma
\label{eq:peaksaccum} \eeq in this manner we estimate that
the last resolved peak has a quantum number $n$ that
satisfies
\beq
 \frac{2n+1}{n^2(n+1)^2} \sim \frac{9 \Gamma}{4m\as^2} = \frac{\Gamma}{E_1}\ll 1
\label{eq:peaksep} \eeq from the values Table~\ref{table1}
and using a width value of the order of 20~MeV there are
only around 5 peaks left before merging.

Because of ISR and beam energy spread, of the order of the GeV for a future
linear collider,
it is unclear whether it could be possible to resolve those peaks
of keV magnitude with this machine. The only potentially detectable peaks
should be the ones belonging to a SQD, provided one has a scenario with widths
of the order of the MeV.

The situation above threshold changes drastically with respect to the ``naive''
Breit--Wigner estimate, as is clearly shown in
Fig.~\ref{fig1} and Fig.~\ref{fig1new} . A few GeV above threshold make
for a factor of 3 of increase compared to the Born cross--section, allowing
a clear distinction between the two cases. Assuming an annual integrated
luminosity of $L_0 = 100$ fb$^{-1}$ and a scale of the extra-dimension 
$R^{-1} =300$ GeV  one finds
around $2.5\times 10^4$ events per year of two quark decay for a center of mass
energy of 10 GeV above threshold (we adopt here the scenario for which the
branching ratio of SQD is essentially 1).
The number of events per year loses an order of magnitude at 
$R^{-1}=700$~GeV, that is about $3 \times 10^3$, as could be inferred
from Fig.~\ref{fig1new}.

As we have already said the decay width of the $KK$ bound
state will be given by twice the decay of the single quark,
as the SQD  dominates, being of the order of up to hundreds
of MeV. For our $\overline{u}_1 u_1$ bound state there are
two possible scenarios of decay
pattern~\cite{Cheng:2002ab}. The first one concerns the
iso-singlet $u_{1_R}$ for which the decay channel into
$W_1$ is forbidden while that into $Z_1$ is heavily
suppressed ${\cal B} (u_{1_R} \to Z_1 u_{0_R})\sim
\sin^2\theta_1 \approx 10^{-2} \div 10^{-3}$ and the
dominant channel is given by $u_{1_R} \to u_{0_R}
\gamma_1$, with ${\cal B}(u_{1_R} \to u_{0_R}
\gamma_1)\approx 0.98$ whose signature is a monochromatic
quark and missing energy of the $KK$ photon, the latter
being the LKP~\cite{Cheng:2002ab}.

For the iso-doublet $u_{1_L}$ the situation is more interesting, as more
channels are available~\cite{Cheng:2002ab}, notably $u_{1_L} \to d_{0_L} W_1$
with ${\cal B}(u_{1_L} \to d_{0_L} W_1) \approx 0.65$ and
$u_{1_L} \to u_{0_L} Z_1$, with ${\cal B}(u_{1_L} \to u_{0_L} Z_1) \approx 0.33$
while the branching ratio into $\gamma_1$ is
negligible ${\cal B}(u_{1_L} \to u_{0_L} \gamma_1)\sim 0.02$.
The decay chain into $W_1$ can follow the scheme:
$ u_{1_L} \to d_{0_L} W_1 \to d_{0_L} \ell_0 \nu_{1_L} \to  d_{0_L} \ell_0 \nu_0 \gamma_1$
with branching ratio given by:
\begin{eqnarray}
{\cal B}(u_{1_L}\to d_{0_L} \ell_0 \nu_0 \gamma_1)
&\approx& {\cal B}(u_{1_L} \to d_{0_L} W_1) {\cal B}(W_1 \to l_0 \nu_1)\cr
&\phantom{\approx}& \times{\cal B}( \nu_1 \to \nu_0\gamma_1) \cr
&\approx& 0.65\times 1/6 \times 1 \approx 10^{-1}
\end{eqnarray}
and alternatively, the same final state could be reached by the scheme:
$ u_{1_L} \to d_{0_L} W_1 \to d_{0_L} \ell_1 \nu_{0_L} \to
d_{0_L} \ell_0 \nu_{0_L} \gamma_1 $. As compared to the iso-singlet case, the
result is a monochromatic quark, a \emph{lepton}  and missing energy in both cases.

The decay into the $Z_1$ channel is $ u_{1_L} \to u_{0_L}
Z_1 \to u_{0_L} \ell_0 \ell_1 \to u_{0_L} \ell_0 \ell_0
\gamma_1$, resulting in a monochromatic quark, two leptons
and missing energy. The branching ratio of the above chain
is:
\begin{eqnarray}
{\cal B}(u_{1_L}\to u_{0_L} \ell_0 \ell_0 \gamma_1)
&\approx& {\cal B}(u_{1_L} \to u_{0_L} Z_1) {\cal B}(Z_1
\to L_0 L_1)\cr &\phantom{\approx}& \times{\cal B}( L_1 \to
\ell_0\gamma_1) \cr &\approx& \frac{1}{3}\times \frac{1}{6}
\times 1 \approx 5\times 10^{-2}
\end{eqnarray}
These leptonic decays of $u_1$ have much cleaner signatures
than the hadronic ones allowing, in principle, for a better
detection of the signal.

In all cases we emphasize that the observable signal of the
bound state production at the linear collider would be
similar to that of the Born pair production except for the
absolute value of the cross-section.  In particular,
assuming for definiteness a linear collider operating
around  the threshold $\sqrt{s}=2 m + E$ GeV we would have
for $R^{-1}=300$ GeV and $E=10$ GeV, in the case of
an iso-singlet bound state (or Born pair production of
${u_1}_R$):
\begin{equation}
e^+e^- \to 2\, \textrm{jets} + E\!\!\!\! /
\end{equation}
with cross section:
\begin{eqnarray}
\sigma (e^+e^- \to 2\, \textrm{jets} + E\!\!\!\! /) &\approx&
\sigma_{\cal B_{KK}}\times \left[{\cal B}({u_1}_R\to u_0
\gamma_1)\right]^2\cr & \approx& 173\, \text{fb}.
\end{eqnarray}
We note that the $\sigma_{\cal B_{KK}}$ for the iso-singlet $u_1$ has to be computed ex-novo and cannot be read from the values of Fig.~\ref{fig1} since it refers to the iso-doublet $U_1$. The singlet and doublet have, when including radiative corrections, different masses and the corresponding pair production threshold is therefore different. For the values of the scale of the extra-dimension $R^{-1}=300$ GeV the mass of the $u_1$ iso-singlet is $m_{u_1}= 351.75$ GeV (slightly lighter than the iso-doublet) and the corresponding Green Function cross-section at an energy offset of $E=10$ GeV is  $\sigma_{\cal B_{KK}} =181$ fb. 
At an $e^+e^-$ collider this signal has a standard model background from $ZZ$ production with one $Z$ decaying to neutrinos and the other decaying hadronically. The cross section for $ZZ$ boson production is $\approx 244$ fb at an energy offset of $E=10$ GeV from the relative thresholds. 
This provides the following estimate for the SM background at 
$R^{-1}=300$ GeV:
\begin{equation}
\sigma_{SM}(2\,\textrm{jets} +E\!\!\!\!/ )\approx 244\,
\text{fb}\times 0.7 \times 0.2 \approx 34\, \text{fb}
\end{equation}
In the case of an iso-doublet bound state (or Born pair
production of ${U_1}$)  the $W_1$ decay chain gives  the
signal:
\begin{equation}
e^+e^- \to 2\, \textrm{jets} +2 \ell +  E\!\!\!\! /
\end{equation}
with cross section (see Fig.~\ref{fig1}):
\begin{eqnarray}
\sigma(e^+e^- \to 2 \textrm{j} +2 \ell +  E\!\!\!\! /)&=&
\sigma_{{\cal B}_{KK}} \left[{\cal B}(u_{1L}\to d_{0_L} \ell_0 \nu_0 \gamma_1 )\right]^2 \cr
&\approx& 275 \text{fb} \times (10^{-1})^2\cr &=& 2.75\, \text{fb}
\end{eqnarray}
while  the $Z_1$ decay chain gives rise to the signature:
\begin{equation}
e^+e^- \to 2\, \textrm{jets} +4 \ell +  E\!\!\!\! /
\end{equation}
with cross sections:
\begin{eqnarray}
\sigma(e^+e^- \to 2 \textrm{j} +4 \ell +  E\!\!\!\! /)&=&
\sigma_{{\cal B}_{KK}} \left[{\cal B}(u_{1L}\to u_{0_L}
\ell_0 \ell_0 \gamma_1 )\right]^2 \cr
&=& 275\text{fb} \times (5\times 10^{-2})^2\cr &\approx& 0.69\, \text{fb}
\end{eqnarray}
Triple gauge boson production, $WWZ, ZZZ$ at a high energy
linear collider has been studied in
refs.~\cite{Barger:1988sq,Barger:1988fd}. It has been found
that these processes receive a substantial enanchement in
the higgs mass range $200$ GeV $<m_H<600$ GeV particularly
the $ZZZ$ channel. As these processes provide a source of
standard model background for our signal we estimate them
both at a value of $m_h =120$ GeV and at a value of $m_h=200$ GeV 
for which the cross sections are
enhanced. Production of $WWZ$ can for instance give rise to
the signature of $2 \textrm{jets}+ 2 \ell + E\!\!\!\! / $
via leptonic decay of the W gauge bosons and hadronic decay
of the $Z$ boson, while the $ZZZ$ production can produce
$2\textrm{jets}+ 4 \ell + E\!\!\!\! / $ via hadronic decay
of one $Z$ while the others decay leptonically with one of them to a pair of $\tau$ which subsequently decay to $\ell \nu\bar{\nu}$ ($\ell =e,\mu$). Estimates
of the resulting cross sections are found using the CalcHEP~\cite{Pukhov:2004ca} and CompHEP~\cite{Boos:2004kh} software. We have verified agreement
with previous results  given in ref.~\cite{Barger:1988fd}
and  for a Higgs mass of $m_h=200$ GeV we find, for $R^{-1}=300$ GeV at
$\sqrt{s} = 2m_{U_1}+10 \approx 724 $ GeV:
\begin{eqnarray}
\sigma(WWZ)&\approx& 72\,  \text{fb} \\ \sigma(ZZZ)&\approx&
\phantom{7}7\, \text{fb}
\end{eqnarray}
We thus find at $\sqrt{s}= 724$ GeV within the standard
model:
\begin{eqnarray}
\label{sigmabg2} \sigma_{\text{SM}}(2 \textrm{j}+ 2 \ell
+ E\!\!\!\! /) &\approx& 70\, \text{fb}\times (0.1)^2\times
0.7 \approx 0.5\, \text{fb}\cr \sigma_{\text{SM}}(2
\textrm{j}+ 4 \ell + E\!\!\!\! /) &\approx& 7\,
\text{fb}\times(0.3)^2\times 0.7\times (0.17)^2 \cr
&\approx&  1.2\times10^{-2}\,\text{fb}
\end{eqnarray}
\begin{table*}[t]
\begin{ruledtabular}
\begin{tabular}{ccccc}
   $R^{-1}$ (GeV)& $m_{u_1}$ (GeV) &$2\, \textrm{jets} + E\!\!\!\!/$ &$2\, \textrm{jets}
   + 2\, \ell + E\!\!\!\!/ $&
   $2\, \textrm{jets} +4\, \ell+ E\!\!\!\!/$\cr
& (iso-singlet) & & $m_h=120$ GeV (200 GeV) & $m_h=120$ GeV (200 GeV) \cr\hline
    300 & 351.7& 121.8 & 13.1 (12.9)& 8.3 (8.2)\cr
400 & 469.0 &81.1 &8.2 (8.1) &5.6 (5.6) \cr
500 & 586.2 &58.7 &5.7 (5.6) &4.2 (4.2)\cr
600 & 703.5 &44.4 &4.1 (4.1) &3.3 (3.3)\cr
700 & 820.7 &35.3 &3.1 (3.1) &2.7 (2.7)\cr
800 & 938.0 &29.0 &2.4 (2.4) &2.3 (2.3)\cr
900 & 1055.2 &24.1 &1.9 (1.9) &1.9 (1.9)\cr
1000 & 1172.5 &20.8&1.6 (1.6) &1.7 (1.7)
 \end{tabular}
\end{ruledtabular}
\caption{\label{table2}Estimate of the statistical
significance $SS$ as defined in Eq.~\ref{statsig} and corresponding to the annual integrated luminosity $L_0=100$ fb$^{-1}$ for  the three channels discussed in the text as a function of  $R^{-1}$ and $\sqrt{s}=2m_{U_1} +E$, assuming an energy offset of $E=10$ GeV from the threshold. The physical threshold of the $2\, \textrm{jets} + E\!\!\!\!/$ channel is different form that of the other two channels as it refers to the $u_1$ iso-singlet level-1 KK quark whose masses  at various values of $R^{-1}$ are given in column two and can  be compared with the corresponding masses of the $U_1$ state form Fig.~\ref{fig1} and Fig.~\ref{fig1new}. The values of the statistical significance for the two multilepton channels have been computed for two values of the Higgs mass $m_h=120\, (200)$ GeV. Again $\Lambda R = 20$.}
\end{table*}
The $2 \textrm{jets}+ 2 \ell + E\!\!\!\! /$ channel could
be potentially contaminated also from $t\bar{t}$ pair
production cross section which at such high energies is
${\cal O}(300)$ fb~\cite{Weiglein:2004hn}. Assuming the top
quarks to decay with probability one to $Wb$ and then the
$W$ gauge boson decay via the leptonic mode (with ${\cal
B}(W\to \ell \nu_\ell) \approx 0.1$) would mimic the signal
with a cross section $\sigma_{SM}(2 \textrm{jets}+ 2 \ell +
E\!\!\!\! /) \approx 3 fb$. However in this case we expect
$b$-tagging of the hadronic jets. Assuming an efficiency in
$b$-tagging of $60\%$ we would get a contribution of $1.2$
fb  to the $2 \textrm{jets}+ 2 \ell + E\!\!\!\! /$
cross-section which has  to be added to that in
Eq.~\ref{sigmabg2}.  This has been done in the calculation
of the statistical significance of table~\ref{table2}.

\begin{figure}[t!]
\includegraphics*[scale=0.745]{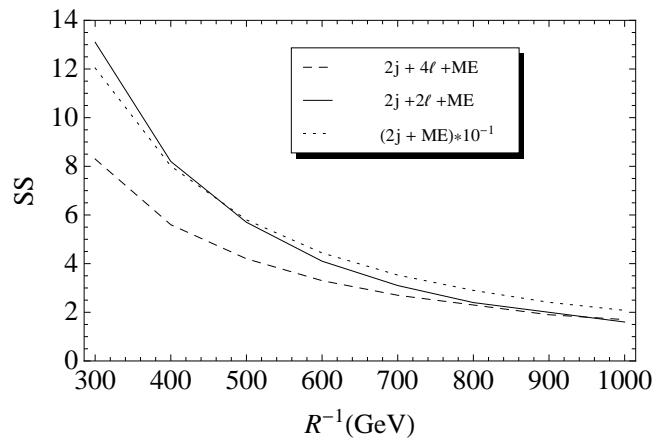}
\caption{\label{fig5} Statistical significance ($SS$) for the various channels discussed in the text as a function of the scale $R^{-1}$ of the extra-dimensions. Note that the statistical significance of the $2j +E\!\!\! /$\ is scaled down by a factor of 10.}
\end{figure}

We conclude providing an estimate of the statistical
significance: 
\begin{equation}
\label{statsig}
SS= \frac{N_{\text{s}}}{\sqrt{N_{\text{s}}+N_{\text{b}}}},
\end{equation}
of the three signals discussed above as related to an
integrated luminosity of $L_0=100$ fb$^{-1}$ ($N_\text{s}$ is the number of signal events and $N_\text{b}$ is the number of background events). These
estimates are given in table~\ref{table2} and Fig.~\ref{fig5}.  Albeit quite
encouraging (especially so the $SS$ of the $2 \textrm{jets} + E\!\!\!\!
/ $) we should bear in mind that the actual observation of
these signals might be not be so easy from the experimental
point of view. Indeed it is quite likely that in a
framework of a quasi degenerate KK mass spectrum the jets
will be typically quite soft and therefore difficult to detect.
It is therefore customary to concentrate on the much
cleaner multilepton signatures~\cite{Cheng:2002ab,Battaglia:2005zf}. 
Indeed a similar analysis to the one given here, but with a
perspective on signals arising at the Compact Linear Collider (CLIC),  regarding the (Born) pair-production of level-1 KK-\emph{leptons} and level-1 KK-\emph{quarks} is given in ref.~\cite{Battaglia:2005zf}.

It is also well known that jets, multilepton and missing energy signals are as well typical of supersymmetric models. Indeed detailed studies have already appeared in the literature regarding the possibility of distinguishing supersymmetric and universal extra dimension models at both the  large hadron collider and linear collider: see for example ref.~\cite{Datta:2005zs,Smillie:2005ar,Freitas:2007rh}.

However, in all cases, angular, invariant mass and/or missing energy distributions of the discussed signals would be identical to those obtained in the Born pair production. In our opinion further detailed analysis of the signals and of the possible SM backgrounds (\emph{and/or competing SUSY signals}) goes beyond the scope of this study, whose main objective is to emphasize the dramatic increase of the bound state cross-section relative to the Born pair production.

\section{Conclusions}
\label{sec:conclusions} Within a universal extra-dimensional model we have considered the formation and decay of a bound state of level-1 quark Kaluza-Klein excitation and its consequent detection at a linear $e^+e^-$ collider. Since $m_{KK}$ should be larger than at least 300~GeV we have used a model with a Coulombic potential. Admittedly this is a model assumption and it should be kept in mind that our results are strictly valid only within this premise, which however has the advantage of providing full analytical expressions for the effect. Being a bound state we have used the Green function technique for the evaluation of its formation cross--section in the threshold region, which is more appropriate than the standard Breit--Wigner picture as it takes into account the binding energy and the peaks of the higher level excitations that coalesce towards the threshold point. The net effect is a dramatic increase of the cross--section in the continuum region right of the threshold. This multiplicative factor is roughly 2.7 for $R^{-1}=300 $ GeV and drops down to 2.2 at $R^{-1}=1000$ GeV. The Green function cross-section would allow more than $\approx 10^4$ events per year even at $R^{-1}=400$ GeV ($m_{U_1}\approx 478$~GeV) for a suitable integrated luminosity of the $e^+e^-$ linear collider ($L_0 =100$ fb$^{-1}$). The number of events $R^{-1}=1000$ GeV ($m_{U_1}\approx 1200$~GeV) would still be $\approx 10^3$ at the same integrated luminosity.

The large difference among the two descriptions of the cross--section should also possibly help in the determination of the correct model for such a heavy bound state outside the SM.

Our analysis of the backgrounds to the final states signals, though very simplified, indicates that the multi-lepton channels have a good statistical significance ($SS \gtrsim 2$) at least up to $R^{-1} =600 \sim 700$ GeV,  which certainly warrants further detailed and dedicated studies of these channels and their backgrounds.  The potentially large (one order of magnitude) estimated statistical significance of the $2j +E\!\!\!\! / $\ channel must be taken however  with great caution because this type of signal may prove difficult to observe as it will be characterized by soft jets within the relatively degenerate mass spectrum of the extra-dimensional model. Further detailed studies are also needed for this channel.

\begin{acknowledgments}
The work of N.~F. was supported by the {\scshape Fondazione Cassa di Risparmio
di Spoleto}.
\end{acknowledgments}

\end{document}